\documentclass[%
reprint,
 amsmath,amssymb,
 longbibliography,
 aps,
prl,
floatfix,
]{revtex4-1}

\usepackage{amssymb}
\usepackage[english]{babel}
\usepackage{graphicx}
\usepackage{multirow}
\usepackage[normalem]{ulem}

\usepackage{dcolumn}
\usepackage{color}
\usepackage{bm}
\usepackage{url} 
\usepackage{hyperref}
\hypersetup{
 colorlinks   = true,
 linkcolor=blue,
  citecolor    = blue,
  urlcolor=blue,
  } 
\usepackage{natbib}
\usepackage{filecontents}

\newcommand{\mycom}{\textcolor{black}}

\AtBeginDocument{%
    \newwrite\bibnotes
    \def\bibnotesext{Notes.bib}
    \immediate\openout\bibnotes=\jobname\bibnotesext
    \immediate\write\bibnotes{@CONTROL{REVTEX41Control}}
    \immediate\write\bibnotes{@CONTROL{%
    apsrev41Control,author="08",editor="1",pages="1",title="0",year="1"}}
     \if@filesw
     \immediate\write\@auxout{\string\citation{apsrev41Control}}%
    \fi
}%
      
\begin{document}

\title{Thermal relaxation in metal films \mycom{limited} by diffuson lattice excitations of amorphous substrates}

\author{Elmira\,M.\,Baeva}
\affiliation{Moscow Pedagogical State University, 29 Malaya Pirogovskaya Street, Moscow, Russia}
\affiliation{National Research University Higher School of Economics, 20 Myasnitskaya Street, Moscow, Russia}
\author{Nadezhda\,A.\,Titova}
\affiliation{Moscow Pedagogical State University, 29 Malaya Pirogovskaya Street, Moscow, Russia}
\author{ Louis\,Veyrat}
\affiliation{Universit\'{e} Grenoble Alpes, CNRS, Grenoble INP, Institut N\'{e}el, 38000 Grenoble, France}
\author{ Benjamin\,Sac\'ep\'e}
\affiliation{Universit\'{e} Grenoble Alpes, CNRS, Grenoble INP, Institut N\'{e}el, 38000 Grenoble, France}
\author{Alexander\,V.\,Semenov}
\affiliation{Moscow Pedagogical State University, 29 Malaya Pirogovskaya Street, Moscow, Russia}
\affiliation{Moscow Institute of Physics and Technology, 9 Institutsky Lane, Dolgoprudny, Russia}
\author{Gregory\,N.\,Goltsman}
\affiliation{Moscow Pedagogical State University, 29 Malaya Pirogovskaya Street, Moscow, Russia}
\affiliation{National Research University Higher School of Economics, 20 Myasnitskaya Street, Moscow, Russia}
\author{Anna\,I.\,Kardakova} 
\affiliation{Moscow Pedagogical State University, 29 Malaya Pirogovskaya Street, Moscow, Russia}
\affiliation{National Research University Higher School of Economics, 20 Myasnitskaya Street, Moscow, Russia}
\author{Vadim.\,S.\,Khrapai}
\affiliation{National Research University Higher School of Economics, 20 Myasnitskaya Street, Moscow, Russia}
\affiliation{Institute of Solid State Physics, Russian Academy of Sciences, Chernogolovka, Russia} 
\begin{abstract}
Here we examine the role of the silicon-based amorphous insulating substrate in the thermal relaxation in thin NbN, InO$_x$, and Au/Ni films at temperatures above 5\,K. The studied samples are made up of metal bridges on an amorphous insulating layer lying on or suspended above a crystalline substrate. Noise thermometry was used to measure the electron temperature $T_e$ of the films as a function of Joule power per unit of area $P_{2D}$. In all samples, we observe the dependence $P_{2D}\propto T_e^n$ with the exponent $n\simeq 2$, which is inconsistent with both electron-phonon coupling and Kapitza thermal resistance. In suspended samples, the functional dependence of $P_{2D}(T_e)$ on the length of the amorphous insulating layer is consistent with the linear $T$-dependence of the thermal conductivity, which is related to lattice excitations (diffusons) for the phonon mean free path smaller than the dominant phonon wavelength. Our findings are important for understanding the operation of devices embedded in
amorphous dielectrics.
\end{abstract}
\maketitle

\section{I. Introduction}

Dielectric substrates covered with an amorphous insulator are often used as a platform for planar electronic devices. The use of an amorphous insulating layer (AIL) in electronic circuits provides technological advantages for optical circuits~\cite{Pernice2012,Kovalyuk2013,Gourgues2019} or electrostatically gated devices~\cite{Novoselov666, Freitag2009} or fabrication of amorphous metallic films used in single-photon detectors~\cite{Smirnov2018,Marsili2013}, calorimeters~\cite{Giazotto2006}, and superconducting resonators~\cite{Lindstrm2009, Dupr2017}. In the latter case, materials have greater robustness with respect to the structural defects. An AIL further ensures a greater yield of practical devices and uniformity of their characteristics~\cite{Steinhauer2020}. Substrates with an AIL are also commonly used by default for the fabrication of strongly disordered  superconducting films, which are of interest for the study of superconductor-insulator transition~\cite{Goldman1998,Sacp2020}. Despite the obvious technological advantages, amorphous materials also cause some undesirable effects in device operations. For instance, in resonators and qubits, two-level systems located in amorphous media lead to energy dissipation and decoherence at low temperatures~\cite{Mller2019}. In electromagnetic detectors, enclosed in an amorphous dielectric, additional phonon bottlenecks in thermal relaxation are observed~\cite{Cherednichenko2007, Sidorova2018,Baeva2018}, which restrict the response timing of the devices. This is manifested in the fact that the magnitude of the phonon escape time $\tau_{esc}$ is much greater than the typical ballistic phonon time of flight $d/v_s$~\cite{Kaplan1979}, where $v_s$ and $d$ are the sound velocity and metal-film thickness, respectively. The observation is now interpreted in terms of strong reflection anisotropy of ballistic phonon at the interface depending on the angle of incidence~\cite{Eisenmenger1976,Sidorova2018,Bezuglyj2018}.

To understand the thermal behavior of thin-film devices, we conduct a systematic study of the electronic heat flow rate, which is proportional to the power-law dependence $P_{2D} \propto T_e^n - T_{b}^n$, where $T_{e}$ and $T_{b}$ are the temperatures of electrons and the bath. Here, the exponent $n$ provides information on the dominant energy-relaxation mechanism. At low temperature $T$, electron-phonon coupling usually mediates the heat flow rate, and in this case, $n$ ranges from $4$ to $6$~\cite{Giazotto2006}. However, in devices on amorphous dielectric substrates, sometimes smaller values of $n$ are observed at low $T$ ~\cite{denisov2018strategy}. This fact cannot be captured by conventional models for electron-phonon coupling.

Heat transfer in amorphous systems is fundamentally different from that in crystalline materials. Above all, the phonon propagation in disordered media is mainly diffusive, since the phonon mean free path ($l_{ph}$) is limited by scattering at defects, boundaries, or two-level systems~\cite{Phillips1987, Beltukov2013}. In amorphous insulators, $l_{ph}$, which defines the thermal conductivity, is dependent on the phonon energy and strongly decreases with increase of $T$~\cite{Smith1978}. On reaching the Ioffe-Regel threshold, which corresponds to  $l_{ph} \approx \lambda/2$, where $\lambda$ is the dominant thermal phonon wavelength, phonons fade quickly, and the heat transfer is implemented by other vibrational excitations, known as diffusons~\cite{Allen1999, Xu2009, Beltukov2013, Ando2018}. As a result, even a thin layer of an amorphous insulator between the metal film and the crystalline substrate can play a crucial role in the thermal transport in real devices. However, this effect is usually not taken into account in modern electrothermal models~\cite{Yang2007}, which \mycom{successfully describe single-photon detectors on polycrystalline materials~\cite{Marsili2011} but poorly apply to devices embedded in an amorphous insulator~\cite{Allmaras2018,Baghdadi}.}

\begin{figure}[h!]
    \includegraphics[scale=1]{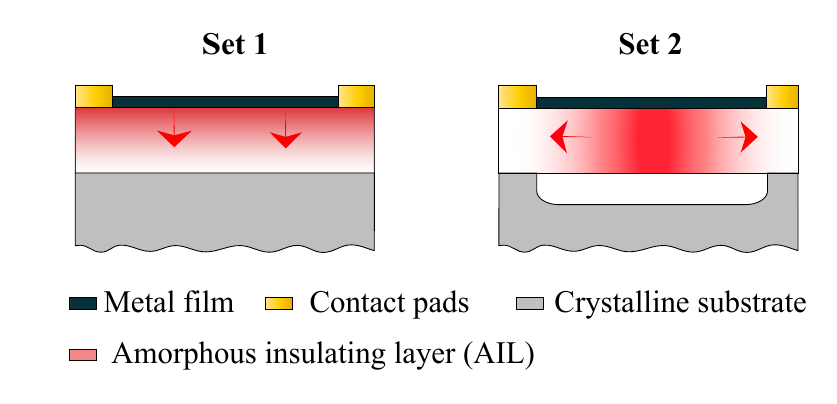}
    \caption{\label{fig_00} Sample configurations. The gradient of red represents the distribution of $T_{ph}$, and red arrows show the direction of heat flow. Set\,1 (left): the metal film is deposited on the amorphous insulating layer. Set\,2 (right): the metal film and the amorphous insulating layer are suspended relative to the  bulk substrate.}
\end{figure}

Here we focus on the study of the role of a silicon-based AIL in thermal transport in thin-film devices at low temperatures. \mycom{Ballistic and diffusive regimes of phonon transport are expected to produce two different scenarios of heat dissipation from a metal film into the substrate.} In the ballistic regime, when the scattering happens only at the interface, we deal with the thermal interface resistance, which corresponds to a $Z_K\propto T_{ph}^{-3}$ dependence~\cite{Swartz1989}, where $Z_K$ is the Kapitza resistance and $T_{ph}$ is the phonon temperature of the metal film. In the diffusive regime, the thermal resistance is determined by the AIL and can be expressed as $Z(T)\propto l_{AIL}/\kappa_{AIL}(T)$, where $\kappa_{AIL}$ is the thermal conductivity of the AIL and $l_{AIL}$ is the effective length of the amorphous insulator. In that case, $Z(T)$ is determined by the temperature dependence of $\kappa_{AIL}(T)$. To discriminate between these two possibilities, we prepare two different sets of samples, deposited on an AIL (see \autoref{fig_00}). In set\,1, we vary the film material and film thickness. In set\,2, a thin metal film is deposited on the AIL, which is suspended above the crystalline substrate. The latter configuration allows us to vary $l_{AIL}$ without changing the material batch. Our main result is that the heat dissipation in both sets \mycom{is controlled by the thickness of the AIL, $l_{AIL}$, and is described by the $Z(T)\propto T^{-1}$ dependence. The data are quantitatively consistent with $\kappa_{AIL}(T)\propto T$ observed in independent calorimetric measurements~\cite{Smith1978,Cahill1987}}. Our finding also demonstrates the insignificant role of the Kapitza resistance at $T>5$\,K.

\section{II. Devices and Methods}

Set\,1 includes films of various metals deposited on commercial substrates based on an amorphous silicon dioxide (SiO$_2$) layer on the top of silicon (Si). The thickness of SiO$_2$, considered here as the AIL, is about $300$\,nm, and the thickness of the Si substrate is 400\,$\mu$m. Set\,2 includes a thin NbN metallic layer deposited on a multilayer (SiN$_x$:H/SiO$_2$/GaAs) substrate. The thickness of SiN$_x$:H/SiO$_2$, considered here as the AIL, is about $700$\,nm. See Appendix A for details. Although the substrates come from different suppliers, we see no difference in the experimental results.

To study the heat dissipation law, we investigate a change of $T_e$ as a function of Joule power per unit area of the metal film ($P_{2D}$). To measure $T_e$ we use resistive thermometry for the Au/Ni sample and noise thermometry for all other samples. For resistive thermometry, we measure the current-voltage $(I-V)$ characteristic, and define $T_e$ from the resistance $R(T_e)=V/I$ and the \mycom{$R(T)$ dependence obtained at low bias current.} For noise thermometry, we measure the current noise spectral density $S_I$ in the current-biased regime and determine the noise temperature as $T_N =S_I \left( dV/dI\right)/4k_B$. The length of our samples ($L$) is chosen to be much longer than the electron-phonon length $l_{e-ph}$ (see Appendix B), which allows us to ignore the electronic contribution to the heat outflow. The configuration also implies uniformity of $T_e$ along the length $L$ in set\,1, and, as a consequence, $T_N = T_e$. The situation in set\,2 is different. Since the temperature gradient is established along the length $L$ in the suspended bridges (see \autoref{fig_00}), $T_N$ reflects the length-averaged temperature.

\mycom{It is instructive to estimate the approximate size of the effective thermal conductances $G\equiv dP_{2D}/dT_e$ at $T=5$\,K expected for different cooling mechanisms. The Joule heat can be removed from the metal film by thermal conduction and thermal radiation. For set\,1, the thermal conduction occurs through the metal film, the AIL, and the bath, connected in series. Here the total $G$ can be defined by the smallest contribution among the electron-phonon conductance $G\approx (0.74- 30)\times 10^{5}$\,WK$^{-1}$m$^{-2}$ (see Appendix B), the Kapitza thermal conductance $G \approx (1.15-1.5)\times 10^{5}$\,WK$^{-1}$m$^{-2}$ (see Appendix C), and the thermal conductance due to diffuson lattice excitations of the substrate. The latter thermal conductivity can be calculated as $G=\kappa_{AIL}/l_{AIL}\approx 3\times 10^{5}$\,WK$^{-1}$m$^{-2}$ for $l_{AIL}=300$\,nm and $\kappa_{AIL}\approx 0.09$\,W/Km for amorphous silica~\cite{Smith1978}. At 5\,K all terms are of the same order of magnitude. At increasing $T$, the thermal relaxation will be governed by the term with the weakest temperature dependence. As we show below, it is the thermal conductance of the AIL that determines $G\propto T$ in the range from 5 to 70\,K. For set\,2, the thermal conduction occurs from the metal film and the AIL to the metal contacts, which represent the thermal bath in this case. Here the thermal conductance is given by $G\approx \kappa_{x}/\alpha L^2$, where $\alpha=\pi^2/64$. In the present experiment $\kappa_x = \kappa d+\kappa_{AIL}d_{AIL}$, where $\kappa$ is the thermal conductivity of the film and $d_{AIL}$ is the thickness of the AIL (see Sec.~III for details). In that case, the total $G$ is determined by the largest contribution to the two-dimensional (2D) thermal conductivity $\kappa_x$ related to electrons and phonons in the film or phonons in the AIL. A preliminary estimate of the thermal conductance due to the AIL is $G\approx 180$\,WK$^{-1}$m$^{-2}$ for $l_{AIL}=50$\,$\mu$m. To estimate $G$ of the film, we suppose that electron and phonon contributions to $\kappa_x$ are of the same order. The thermal conductance of electrons determined according to the Wiedemann–Franz law yields $G\approx 0.4$\,WK$^{-1}$m$^{-2}$ for $L=50$\,$\mu$m. Thus, in suspended devices, due to $d_{AIL}/d \gg 1$, the thermal conductance of the amorphous substrate is expected to dominate already at T=5\,K. Finally, electromagnetic thermal radiation provides the least-effective channel for cooling ($G=4\sigma T^3\approx 2.8\times 10^{-5}$\,WK$^{-1}$m$^{-2}$ at 5\,K according to the Stefan–Boltzmann law, where $\sigma$ is the Stefan–Boltzmann constant), and we do not consider it further.}

\section{III. Results}
\subsection{Set 1}\label{section1}
 Here we investigate metal films based on various materials (NbN, InO$_x$, and Au/Ni) deposited on similar SiO$_2$/Si substrates. Images of the samples and schematic sketches of the thermometry methods are presented in~\autoref{fig_01}(a-c): see Appendix A for details. The materials substantially differ in the level of disorder, the resistivity varies by $3$ orders of magnitude, and the film thickness $d$ varies in the range from 5 to 130\,nm. \autoref{fig_01}(d) shows the measured $T_e$ as a function of $P_{2D}$ on a log-log scale. In the limit of small heating, $T_e$ remains close to the bath temperature $T_b$, which corresponds to the plateau in the $T_N(P_{2D})$-dependence. Note that $T_b$ is sample specific. For superconducting NbN and InO$_x$ films, $T_b$ is chosen to be slightly higher than the critical temperature of the superconducting transition to keep samples in the normal state. For the nonsuperconducting Au/Ni device, $T_b=0.5$\,K; however, $T<15$\,K data were omitted due to the lack of accuracy of resistive thermometry in this temperature range. With intense heating, when $T_e\gg T_b$ regime is achieved, we observe the same dependency for all samples characterized by the power-law  dependence $P_{2D}\propto T_e^2$ (the dashed guideline). Overall, the measured $T_e(P_{2D})$-dependence can be fitted by the equation $P_{2D}=\Sigma^{2D} (T_e^n-T_b^n)$ (solid lines), where $\Sigma^{2D}$ is the two dimensional cooling rate and $n=2$ is the exponent in heat-outflow law. The fact that a similar dependence is observed for different materials clearly indicates that the observed effect is mediated by the substrate.
\begin{figure}[h!]
    \includegraphics[scale=1]{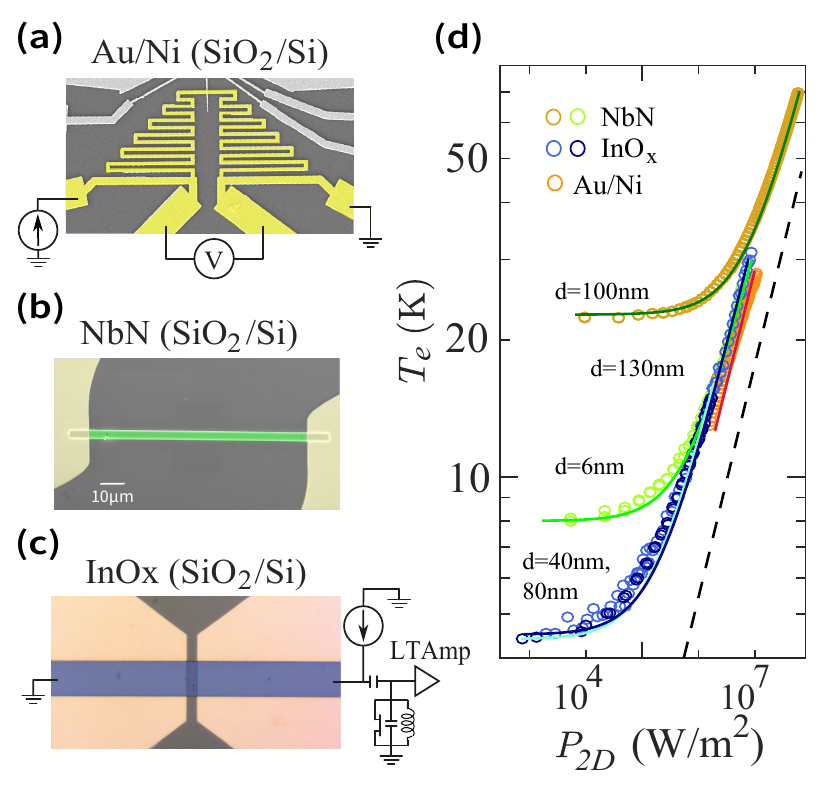}
    \caption{\label{fig_01} Devices and the experimental methods. The photographs from scanning electron and optical microscopes of a few samples with schematic sketches of the experiment: resistive thermometry, applied for the Au/Ni device (a), and noise thermometry, utilized for NbN device (b) and InO$_x$ device (c). (d) The measured $T_e$ is presented as a function of $P_{2D}$ \mycom{on a log-log scale for samples with various metal-film thicknesses $d$}. The solid lines represent the fits, obtained with the expression \mycom{$P_{2D} =\Sigma^{2D} (T_e^{n}-T_{b}^n)$} with the power index $n=2$. The dashed black line displays $P_{2D}\propto T_e^2$ as a guide for the eye.}
\end{figure}

\begin{figure}[h!]
    \includegraphics[scale=1]{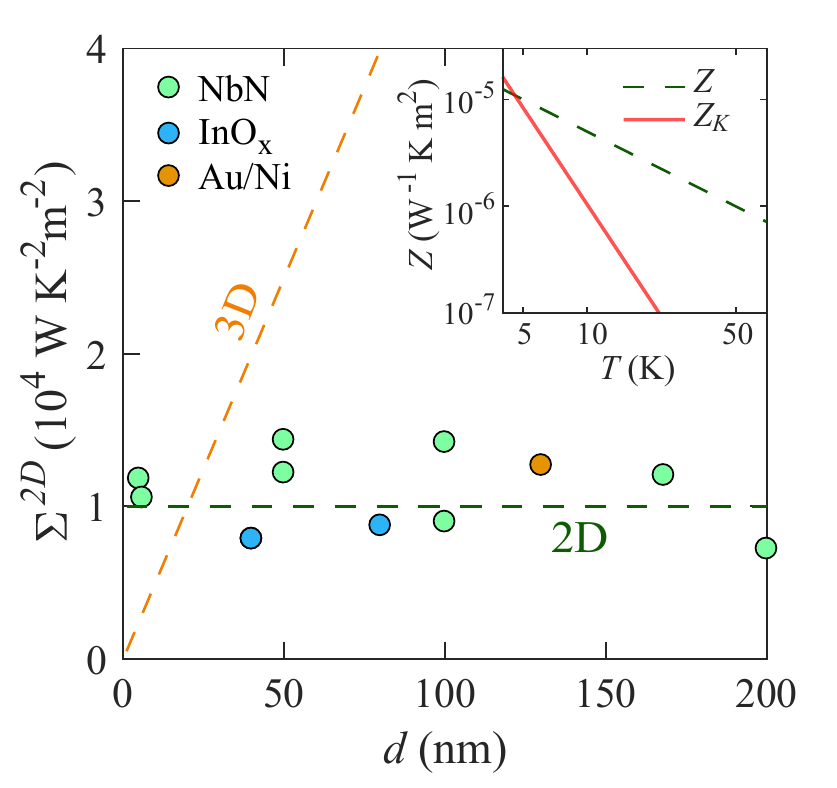}
    \caption{\label{fig_02} The thickness dependencies of the experimental 2D cooling rate $\Sigma^{2D}$. The symbols represent $\Sigma^{2D}$ for different materials (NbN, InO$_x$, Au/Ni) deposited on a SiO$_2$/Si substrate. The dashed green and orange lines highlight 2D and 3D heat dissipation. The inset shows the temperature dependencies of the thermal resistance $Z$ and the Kapitza resistance $Z_K$ (red line) on a log-log scale.}
\end{figure}

In \autoref{fig_02} we plot the fit parameter $\Sigma^{2D}$ versus the film thickness $d$. Here we add the series of data for NbN samples with $d$ varying by $2$ orders of magnitude, each symbol on a graph corresponding to an individual sample. $\Sigma^{2D}$ remains insensitive to $d$ and is close to $\Sigma^{2D}_{av}=10^4$\,W K$^{-2}$m$^{-2}$ for all samples studied (the dashed green line). In addition, we compare our data with three-dimensional (3D) heat dissipation, which is defined as $IV/\Omega = \Sigma_{3D} (T_e^n - T_b^n)$, where $\Omega$ is the sample volume and $\Sigma_{3D}\approx \mathrm{const}$. The dashed orange line shows the trend expected for 3D heat dissipation, which does not occur here. Summarizing the experimental results from \autoref{fig_01} and \autoref{fig_02} our main findings are $P_{2D}\propto T_e^2$ and $\Sigma^{2D} \approx \mathrm{const}$. The temperature dependence of $P_{2D}$ we find is in contradiction with the model of thermal boundary resistance, which implies that the heat flow is due to ballistic phonons and predicts $P_{2D}\propto T_{ph}^4$. In the inset \autoref{fig_02}, we compare typical experimental values of the thermal resistance, defined as $Z=1/ n \Sigma^{2D}_{av} T^{n-1}$, with the expected Kapitza thermal resistance $Z_K$. Here $Z_K$ is calculated in the frame of the acoustic mismatch model (AMM)~\cite{Swartz1989}. The red line shows estimations for the interfaces between all metallic films and the SiO$_2$ layer (see Appendix C for details). From the graph, we find $Z> Z_K$ at $T>5$\,K, which indicates a gross violation of ballistic phonon flight in this temperature range. This observation demonstrates that in this temperature range the probability of a phonon backscattering approaches unity. 

\begin{figure}[h!]
    \includegraphics[scale=1]{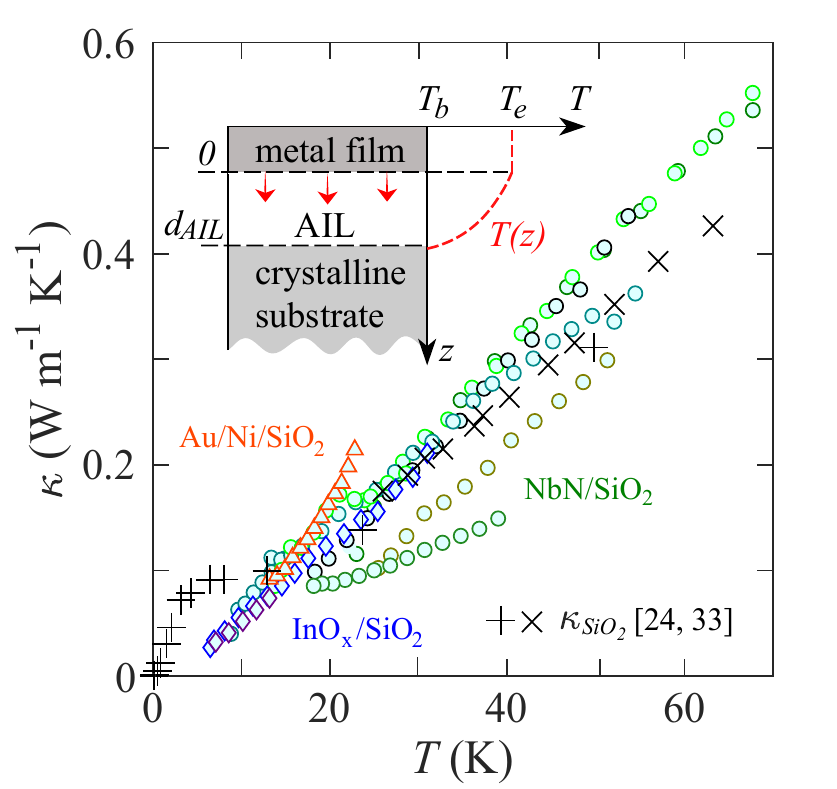}
    \caption{\label{fig_03} 
Comparison of thermal conductivity derived from differential thermal resistance with the data from calorimetric measurements. The colored symbols correspond to the thermal conductivity of the AIL calculated with Eq.\eqref{eq_kox} for NbN/SiO$_2$, InO$_x$/SiO$_2$, and Au/Ni/SiO$_2$ devices. The black symbols represent the data for SiO$_2$ thermal conductivity ($\kappa_{SiO_2}$) obtained with calorimetric measurements~\cite{Smith1978,Cahill1987}. The inset shows the thermal relaxation in a metal film deposited on the substrate with the AIL leads to a temperature gradient across the AIL in the $z$-direction.}
\end{figure}

To gain further understanding of the $P_{2D}\propto T_e^2$ dependence we consider that the measured heat-dissipation dependence is restricted by the diffusive heat propagation in the AIL. It is convenient to introduce a gradient of the phonon temperature $T_{ph}$ across the AIL (see the inset in \autoref{fig_03}). The boundary conditions at the upper and bottom surfaces of the AIL are $T_{ph}(z = 0)=T_e$ and $T_{ph}(z = d_{AIL})\simeq T_{b}$. Here we ignore the contribution of the crystalline Si substrate since its thermal resistance is $3$ orders of magnitude lower than the thermal resistance of the AIL (see Appendix E for details). The heat-flux continuity condition is given by the expression 
\begin{equation}
\label{eq_sub}
\frac{d}{dz}\left(\kappa_{AIL} \frac{dT_{ph}(z)}{dz}\right)=0
\end{equation}
where $\kappa_{AIL}$ is the thermal conductivity of the AIL. 
The solution of Eq.\eqref{eq_sub}, described in Appendix D, connects $\kappa_{AIL}(T_e)$ and the thermal resistance $Z\equiv dT_e/d P_{2D}$. Note that $\kappa_{AIL}$ also depends on $T_{ph}$ at the $z$-axis coordinate. Thus, under the assumption of a substrate effect, one obtains the temperature dependence of $\kappa_{AIL}$:
\begin{equation}
\label{eq_kox}
\kappa_{AIL}(T_e)= \frac{d_{AIL}}{Z}
\end{equation}

The differential thermal resistance $Z$ yields a nearly linear temperature dependence of $\kappa_{AIL}$ under the assumption of the substrate effect.
\autoref{fig_03} shows $\kappa_{AIL}$ obtained via Eq.\eqref{eq_kox} as a function of $T$ on a linear scale. Colored symbols correspond to $\kappa_{AIL}$ obtained for all samples. For comparison, we plot the values of the thermal conductivity of SiO$_2$ ($\kappa_{SiO_2}$) obtained by calorimetry~\cite{Smith1978,Cahill1987}. The measured $\kappa_{AIL}$ is in good agreement with the calorimetric data $\kappa_{SiO_2}$ over the temperature range from $20$ to $70$\,K. In the range from $4$ to $20$\,K our estimation of $\kappa_{AIL}$ deviates from $\kappa_{SiO_2}$, and the origin of the discrepancy is unclear. In the next section, we verify the functional dependence of $Z$  on the length of the AIL in suspended devices.

\subsection{Set 2}\label{section2}

\begin{figure}[h!]
    \includegraphics[scale=1]{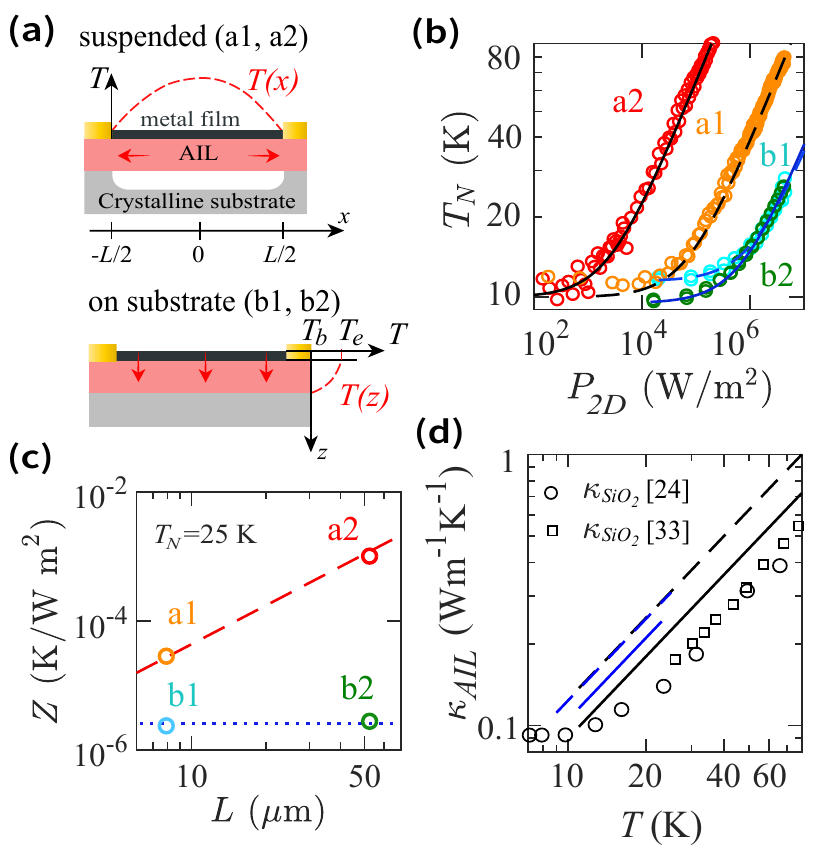}
    \caption{\label{fig_04} Comparison of heat transport in the suspended NbN and NbN devices on the substrate. (a) Schematics of the relevant mechanisms of the thermal transport of suspended samples~(a1 and a2) and samples on the substrate~(b1 and b2). (b) $T_N$ as a function of $P_{2D}$ on a log-log scale at the bath temperature $T_{b} \approx 10$\,K. The data for samples a1 and a2 are displayed with orange and red symbols, while the data for samples b1 and b2 are displayed with blue and green symbols. The solid and dashed black lines represent the fits of the data for samples a1 and a2 with $T_N$~(see the main text). The solid and dashed blue lines represent the fits of the data for samples b1 and b2 with $T_N=\sqrt{(P_{2D})/\Sigma^{2D}+T_{b}^2}$. (c) The thermal resistance $Z$ versus length of the samples $L$ at $T_{N} = 25$\,K \mycom{on a log-log scale}. (d) The calculated thermal conductivity of the AIL for the studied NbN samples in comparison with $\kappa_{SiO_2}$. Solid and dashed black lines correspond to the values of $\kappa_{AIL}$, estimated with Eq.~\eqref{eq_kx}, for samples a1 and a2. The blue lines correspond to $\kappa_{AIL}$ obtained with Eq.\eqref{eq_kox}, for NbN samples b1 and b2. The symbols demonstrate the data for $\kappa_{SiO_2}$~\cite{Smith1978,Cahill1987}.}
\end{figure}

As follows from Eq.~(\ref{eq_kox}), for unequivocal identification of the AIL bottleneck mechanism a variation of the amorphous-layer thickness is desirable. To this end, we perform analogous experiments in metal-on-AIL bridges suspended in a vacuum above the crystalline substrate. The idea is that the Joule heat released in the metal film propagates along the bridge toward the massive Ohmic contacts [see \autoref{fig_04} (a)] so that the thermal gradient is directed along the bridge and its length $L$ controls the solution of the heat-balance equation. Using 5-nm-thick NbN films deposited on 700-nm-thick AIL, we ensure that the thermal relaxation is determined by the thermal properties of the AIL. In this experiment, the AIL consists of two layers of amorphous insulators: 200-nm-thick SiN$_x$:H and 500-nm-thick SiO$_2$ layers. We also fabricate two nonsuspended devices on the same substrate (see  \autoref{table_02} in Appendix A for detailed information). A schematic illustration of sample set\,2 and the thermal gradients is given in \autoref{fig_04} (a) for suspended (a1, a2) and nonsuspended (b1, b2) devices.

Assuming that the phonon temperature is uniform within the cross section of the bridge and $T_e=T_{ph}=T$ we obtain a thermal-balance equation for a suspended device:
\begin{equation}
\label{eq_kx}
\frac{\partial}{\partial x}\left(\kappa_x  \frac{\partial T}{\partial x}\right)=-P_{2D},
\end{equation}
where $\kappa_x$ is the total (two-dimensional) thermal conductivity of the bridge and $x$ is the coordinate along the bridge. In~\autoref{fig_04} (b) we plot the measured noise temperature $T_{N}$ for all four devices from set 2 as a function of $P_{2D}$. We observe the same power-law functional dependence as before, $P_{2D}\propto(T_{N}^2-T_{b}^2)$, with a notable distinction. Here the suspended devices a1 and a2 exhibit strong length dependence: the longer device a2 is much easier to heat up. 
This effect is not observable for devices b1 and b2 lying on the substrate, and it is similar to the data for set\,1 discussed above. The observed length dependence is consistent with the fact that the solution of Eq.~(\ref{eq_kx}) in the reduced coordinate $\tilde{x}\equiv x/L$ is a universal function of length $T^{2}(\tilde{x})\propto L^2P_{2D}$ at $T\gg T_b$. This is illustrated in \autoref{fig_04} (c), which shows the thermal resistance $Z=dT_{N}/dP_{2D}$ evaluated at $T_{N} = 25\,$K as a function of $L$. The data for suspended devices are indeed consistent with $Z\propto L^2$ ( see the dashed line), whereas $Z\approx\mathrm{const}$ is observed for the nonsuspended devices (see the dotted line).

The observation of a parabolic temperature dependence $P_{2D}\propto(T_{N}^2-T_{b}^2)$ is consistent with a linear temperature dependence $\kappa_x\propto T$ of the heat conductivity in Eq.~(\ref{eq_kx}). This yields the estimation of the thermal resistance $Z=\pi^2 L^2/64 \kappa_x$. In the present experiment $\kappa_x = \kappa_{NbN}d_{NbN}+\kappa_{AIL}d_{AIL}$, where $\kappa_{NbN}$ is the thermal conductivities of NbN (electron and lattice). In our devices with $d_{AIL}\gg d_{NbN}$, the contribution of the NbN film is negligible and we obtain $\kappa_x \approx \kappa_{AIL}d_{AIL}$, also consistent with an independent estimate of $\kappa_{NbN}$. Solving Eq.~(\ref{eq_kx}) for suspended devices, we fit the data from \autoref{fig_04} (b) and extract the corresponding $\kappa_{AIL}$. In this two-step procedure we find a temperature profile along the bridge $T^{2}(\tilde{x})=T^{2}_{b}+\left(T_{max}^{2}-T^2_{b}\right) \left(1-4\tilde{x}^2\right)$, where $|\tilde{x}|\leq1/2$, and subsequently evaluate the noise temperature of the device \mycom{$T_{N}=\int T(\tilde{x})d\tilde{x}$}. The data for nonsuspended devices are analysed according to Eq.~(\ref{eq_kox}) in the same manner as set 1 before. \autoref{fig_04} (d) shows the obtained linear dependencies of $\kappa_{AIL}$ on temperature for all four devices, with the same line styles  as the fits in \autoref{fig_04} (b). Note that the temperature ranges, which vary for the samples, correspond to the range of $T_{N}$ variation in the experiment. Because of the lack of data on the calorimetric measurement of the thermal conductivity of amorphous SiN$_x$:H we compare our results with data for SiO$_2$ only. The data are consistent with each other within the uncertainty of about 15\%, as well as with independent calorimetric measurements of $\kappa_{SiO_2}$, shown by symbols. The plotted data of $\kappa_{SiO_2}$ are the same as in \autoref{fig_03} in other temperature range. In the range from $10$ to $60$\,K, our estimation of $\kappa_{AIL}$ for set\,2 slightly deviates from $\kappa_{AIL}$ for set\,1. The origin of the discrepancy can be explained by the uncertainty in the thermal conductivity of SiN$x$:H, which is expected to exceed the thermal conductivity of SiO$_2$.
\section*{IV. Discussion}

We now summarize and discuss our observations of thermal transport in the samples studied. The observed heat flow rate with the exponent $n=2$ is not compatible with the electron-phonon relaxation or the phonon relaxation mediated by the Kapitsa thermal resistance~\cite{Giazotto2006}. Experimental findings such as (i) the similar value of $\Sigma_{2D}$ observed for the different metal films and (ii) the functional dependence of $Z$ on the length of the AIL strongly indicate that the heat transport is mediated by the thermal properties of the AIL. This effect is also confirmed by the obtained dependence $\kappa_{AIL}\propto T$.

Heat transfer in an AIL is different from heat transfer in a crystalline material, and $l_{ph}$ in amorphous solids strongly decreases with increasing $T$. \mycom{In silicon-based materials at} a certain temperature, $l_{ph}$ reaches the Ioffe-Regel criterion. Note that the phonons, considered as plane waves with well-defined wave vector $q$ and frequency $\omega$, are in a strongly scattering regime and cannot propagate and transfer heat here~\cite{Allen1999}. By contrast, the heat transfer is expected to be controlled by diffusons~\cite{Allen1999, Xu2009, Beltukov2013, Ando2018}. In this framework, the thermal conductivity is determined by $\kappa_{AIL} \propto \int  g(\omega )D(\omega )C(\omega/T) d\omega$, where $g(\omega)$ and $D(\omega)$ are the density and diffusivity of vibrational modes, respectively, and $C(\omega/T)= x^2/sinh^2x$ with $x=\hbar\omega/2k_BT$, is the specific heat capacity of a harmonic oscillator. For diffusons, the functions $g(\omega )$ and $D(\omega )$ are approximately constant in some frequency interval~\cite{Xu2009, Beltukov2013}, which leads to $\kappa \propto k_B^2 T/(\hbar a_0)$ in given temperature range, where $a_0$ corresponds to an interatomic distance. In our study, the Ioffe-Regel threshold is expected to occur at approximately 10\,K~\cite{Taraskin2000, Beltukov2013}. The estimated phonon wavelength $\lambda$ and phonon mean free path $l_{ph}(\omega)$ at 10\,K are $9$\,nm and approximately $4.5$\,nm, respectively. However, in our experiment we do not observe the plateau of $\kappa$, which has been found with calorimetric measurements~\cite{Cahill1987} in the range from $5$ to $20$\,K (see Figure~\ref{fig_03}). The position of this plateau in $\kappa(T)$ agrees with the position of the boson peak in vitreous SiO$_2$~\cite{Nakayama2002,Ando2018}, and is usually considered as the boundary between ballistic and diffusely propagating lattice excitations. It is possible that the fact that we do not observe a plateau is related to the fabrication process for contemporary commercial substrate materials, and this remains to be clarified in future experiments.

Our results also shed light on a recent proposal of a universal energy-relaxation bottleneck in thick strongly disordered metallic films~\cite{Baeva2020}. In this work, since $\Sigma_{2D} \approx \mathrm{const}$ in \autoref{fig_02}, we do not observe an impact of phonon scattering in NbN films on heat transfer up to thicknesses on the order of 200\,nm. Further studies of disordered metal films on crystalline substrates are required to verify the strong phonon-scattering effects inherent in disordered metal films. Our results may also be useful for interpreting the thermal transport in devices embedded in an AIL at low temperatures.  At subkelvin temperatures, when the phonon mean free path is long enough, the exponent in the heat flow rate $n$ is determined by the internal properties of the metal film~\cite{Underwood2011}. Thus, the decrease of $n$ at low $T$ in other devices on amorphous dielectric substrates~\cite{denisov2018strategy} may be related to the transition from the electron-phonon coupling regime to the substrate effect. In addition, the AIL may be a possible candidate for phonon-filtering applications~\cite{Melkonyan2003}, since $l_{ph}$ in amorphous insulators is frequency dependent ($l_{ph}\propto \omega^{-4}$~\cite{Beltukov2013, richet2021encyclopedia}).
In this respect, the phonon scattering in amorphous insulators resembles phonon propagation as a low-pass energy filter, and the low-energy modes have an opportunity to propagate, while the high-energy phonons cannot pass. 
The energy filtration can be an intriguing alternative to the models of strong acoustic mismatch, which assume ballistic phonon filtering at the interface depending on the angle of incidence~\cite{Sidorova2018}.

\section*{V. Conclusion}
In conclusion, we conduct a systematic study of the electronic heat flow rate in metal films on silicon-based amorphous insulating substrates at temperatures above 5\,K. For samples lying on the substrate, the observed two-dimensional heat-relaxation ldependence with $P_{2D}=\Sigma^{2D} (T_e^n-T_{b}^n)$ with exponent $n\simeq 2$ is inconsistent with both electron-phonon cooling and Kapitza resistance. For samples suspended above the crystalline substrate, we observe length-dependent heat relaxation with the same exponent. This effect is quantitatively explained by the low thermal conductance of the amorphous insulating layer. The exponent $n\simeq 2$ is related to the well-known linear temperature dependence of thermal conductivity in amorphous solids, which is described by the concept of diffuson lattice excitations. Our findings refine the understanding of thermal transport in mesoscopic devices embedded in an amorphous dielectric.

\section*{Acknowledgements}
\begin{acknowledgements}
We are grateful to E.S. Tikhonov for fruitful discussions and help in the initial stage of this work. We thank A.O. Denisov for sharing experimental results for Au/Ni sample. We are grateful to M. Rocci for fabrication of Au/Ni devices as a part of a different project. The transport and noise measurements were funded by the Russian Science Foundation (Project No. 19-72-10101). The theoretical analysis was supported by the Council on grants of the President of the Russian Federation (No. MK-1308.2019.2). Resistive thermometry was performed under the state task of the ISSP RAS.
\end{acknowledgements}

Note added.—Recently we have become aware of Ref.~\cite{Baggioli2019}, which provides insights about the presence of diffusivelike damping of vibrational excitations in amorphous materials and even ordered crystals, and its strong influence on low-T properties of solids.

\section*{\label{appendix:A} Appendix A: Details of sample fabrication and experimental setup}
The films of set\,1 are deposited on a SiO$_2$/Si substrate obtained from NOVA Electronic Materials LLC. The 280-300-nm SiO$_2$ layer is produced by the thermal oxidation of a crystalline Si substrate.

The Au(120\,nm)/Ni(10\,nm) bilayer is deposited by means of electron-beam lithography and the lift-off technology~\cite{denisov2018strategy}. Here we use the residual resistance ratio, $r_R=\left(R_{300K}-R_{20K}\right)/R_{20K}$, to characterize the metal films. Positive values of $r_R$ correspond to  $dR(T)/dT>0$, and vice versa. The bilayer is characterized by  $r_R=2$ and resistivity $\rho=1.6\times 10^{-8}$\,$\Omega$\,m at 10\,K. 

The NbN films of set 1 are deposited on a substrate at room temperature with dc magnetron sputtering. The NbN films have $r_R=-0.3$, $\rho=10^{-5}$\,$\Omega$\,m at 20\,K, and critical temperature of the superconducting transition $T_c=$13.5\,K in 200-nm-thick film. The Ti(5\,nm)/Au(200\,nm) metal pads to NbN are fabricated with standard photolithography and thermal evaporation.

The amorphous InO$_x$ films are characterized by $r_R=-0.3$, $\rho=8\times 10^{-5}$\,$\Omega$\,m at 4\,K, and $T_c=2.7$\,K. The metal Ti/Au leads are formed on the substrate before evaporation of InO$_x$ films from In$_2$O$_3$ granules at room temperature~\cite{Sacp2015}. 

All films are patterned into a bridge or a meander by a plasma-chemical etching or lift-off. Photographs of samples are presented in \autoref{fig_01}(a), and the sizes and resistance of the samples are presented in (\autoref{table_01}).

\begin{table}[h!]
\caption{\label{table_01} Parameters of set\,1. $d$ is the thickness, $w$ the width, $L$ the length and $R$ the resistance measured above $T_c$ for superconducting NbN and InO$_x$ films or at 10K for Au/Ni bilayer.}
\begin{tabular}{ccccc} 
\hline \hline
Sample & \begin{tabular}[c]{@{}c@{}}$d$ \\ (nm)\end{tabular} & \begin{tabular}[c]{@{}c@{}}$w$\\ ($\mu m$)\end{tabular} & \begin{tabular}[c]{@{}c@{}}$L$\\ ($\mu m$)\end{tabular} & \begin{tabular}[c]{@{}c@{}}$R$\\ ($k\Omega$)\end{tabular} \\ \hline
\multirow{8}{*}{\rotatebox[origin=c]{0}{NbN}} & 5 & 0.66 & 12.3 & 129 \\ \cline{2-5} 
 & 6 & 10.3 & 10.2 & 1.75  \\ \cline{2-5} 
 & \multirow{2}{*}{50} & 0.55&12.6 & 4.23 \\ \cline{3-5} 
 &  & 0.97 & 25.8 & 5.16 \\ \cline{2-5} 
 & \multicolumn{1}{c}{} & 0.54 &12.9 & 2.5  \\ \cline{3-5} 
 & 100 & 0.95 & 25.6 &2.68 \\ \cline{2-5} 
 & 168 & 3.3&93 & 1.44\\ \cline{2-5} 
 & 200 &1&22.5  & 1.76  \\ \hline
\multirow{2}{*}{\rotatebox[origin=c]{0}{InO$_x$}}  & 40 &3.7&3.1 & 1.66 \\ \cline{2-5} 
 & 80 & 10.3&9.2 & 1.02  \\ \hline
\rotatebox[origin=c]{0}{Au/Ni} & 130 &  0.25&105 & 0.03 \\ \hline
\hline
\end{tabular}
\end{table}

The 5-nm NbN film of set\,2 is deposited on a SiN$_x$:H/SiO$_2$/GaAs substrate at 250$^{\circ}$C with dc magnetron sputtering. The SiO$_2$ layer is grown on a crystalline GaAs substrate by chemical vapor deposition (CVD) and the SiN$_x$:H layer is grown by plasma-enhanced chemical vapor deposition (PECVD) at 250$^{\circ}$C~\cite{Garmash2015}. The thicknesses of AIL SiN$_x$:H and SiO$_2$ membranes are 200 and 500\,nm, respectively. The NbN film has $r_R=-0.25$ and resistivity $rho$ of about $=4\times 10^{-6}$\,$\Omega$\,m at 10\,K. $\rho$ of NbN in set\,2 is $2$ times smaller than in set\,1 due to the higher temperature of deposition. The NbN film is patterned to form a bridge-type structure with Ti/Au metal pads. To fabricate the suspended structure (membrane), beyond the NbN bridge, SiN$_x$:H and SiO$_2$ layers are dissolved in hydrofluoric acid, and GaAs under the membrane is etched in a solution of hydrogen peroxide, ammonia, and water.

\begin{table}[h!]
\caption{\label{table_02} Parameters of set\,2. $d$ is the thickness, $w$ the width, $L$ the length and $R$ the resistance measured above $T_c$ for superconducting NbN films.}

\begin{tabular}{ccccc}
\hline
\hline
Sample & $d$ & $w$ & $L$ & $R$ \\ \hline
 & (nm) & ($\mu$m) & ($\mu$m) & (k$\Omega$) \\ \hline
b1 & 5 & 6.5 & 8.7 & 4.1 \\ \hline
b2 & 5 & 6.35 & 50 & 13 \\ \hline
a1 & 5 & 3.1 & 7 & 1.6 \\ \hline
a2 & 5 & 3.77 & 53.8 & 7.2 \\ \hline
\hline
\end{tabular}
\end{table}

The noise and resistance measurements are performed in a homemade $^4$He insert, inside which the samples are in a vacuum. For noise thermometry, the current-noise spectral density is measured with a resonant tank circuit at the input of a homemade low-noise amplifier (LTAMP) with a gain of about $6$\,dB, input current noise of approximately $10^{-27}$\,A$^2$/Hz and dissipated power of approximately $250$\,$\mu$W (see the sketch in~\autoref{fig_01}(c)). The output noise signal of the LTAMP is amplified by a cascade of low-noise amplifiers with a gain of $75$\,dB in total, and then it is passed through a system of band-pass filters and measured by a power detector. The current dependence of the power is averaged over several measurements to reduce random error, and the uncertainty of the measured temperature is within 0.4\,K. Calibration is achieved with equilibrium Johnson-Nyquist noise thermometry. For this purpose, we use a commercial high-electron-mobility transistor connected in parallel with the device, which is depleted otherwise. At high sample resistance (above $1k\Omega$), the setup has a bandwidth $\Delta f $ of approximately $1$\,MHz around a center frequency of $40$\,MHz. The low-quality factor of the resonant tank circuit precludes the use of noise thermometry in the Au/Ni device, which has a resistance of about 30\,$\Omega$. For the Au/Ni device the electron temperature is measured by resistive thermometry as an alternative [see the sketch of setup in~\autoref{fig_01}(a)]. Since the accuracy of resistive thermometry crucially depends on the function $dR/dT$, we leave only data that provide the uncertainty of the measured temperature within 0.15\,K. Our results obtained by resistive thermometry are qualitatively consistent with the data from local noise thermometry in the same material performed in the work reported in Ref.~\cite{denisov2018strategy}. The relaxation rates smaller by a factor of $3-4$ obtained in Ref. \cite{denisov2018strategy} may result from the substrate overheating effect, which was not anticipated in that work.

\section*{\label{ap2} Appendix B: Estimation of the electron-phonon cooling}

The electron-phonon length $l_{e-ph}$ can be calculated as $l_{e-ph}=\sqrt{\mathcal{L}/\rho n \Sigma_{e-ph} T^{n-2}}$ or $l_{e-ph}=\sqrt{\tau_{e-ph} D}$, where $\Sigma_{e-ph}$ and $\tau_{e-ph}$ are the electron-phonon coupling constant and the electron-phonon relaxation time, $n$ is the exponent, $D$ is the diffusion coefficient, and $\mathcal{L}$ is the Lorentz number. Using $\Sigma_{e-ph}=7.7\times 10^9$\,WK$^{-n}$m$^{-3}$, $n=5.07$, for Au~\cite{Saira2020}, $\Sigma_{e-ph}=1.85\times 10^9$\,WK$^{-n}$m$^{-3}$, $n=6$, for InO$_x$~\cite{Ovadia2009}, and the value of $\tau_{e-ph}$ extrapolated to 10\,K ($\tau_{e-ph}=9$\,ps) and $D=0.3$\,cm$^2$/s for NbN~\cite{Gousev1994} we calculate $l_{e-ph}$ is $350$\,nm for Au, $18$ nm for NbN at 10\,K, and $6.6$\,nm for InO$_x$ at 5\,K. The effective thermal conductance $G=dP_{2D}/dT_{e}$ is given by  $G=n\Sigma_{e-ph} T^{n-1} d$ or $G=C_e \tau_{eph}^{-1} d$, where $d$ is the film thickness, $C_e$ is the electron specific heat capacity.

\section*{\label{ap3} Appendix C: Estimation of Kapitza resistance}

The Kapitza resistance in analogy to the Stefan-Boltzmann law is described by the heat-dissipation cooling law $P=\Sigma_{K}A\left(T_{ph}^4-T_{b}^4\right)$, where $\Sigma_{K}$ is the cooling rate due to Kapitza resistance. $\Sigma_{K}$ can be estimated with AMM or diffuse mismatch model (DMM). The AMM describes phonon propagation through the interface between two media in analogy to Snell’s law for electromagnetic waves. The probability of phonon transmission depends on the angle of the incident phonon related to the critical angle, where the critical angle is determined by the acoustic properties of the medium~\cite{Little1959}. In the DMM the diffusive scattering of phonons at the interface is taken into account, and thus the phonon transmission probability depends only on the phonon densities of the states and sound velocities of the two media~\cite{Swartz1989}. In the DMM $\Sigma_{K}$ between the film phonons and the substrate phonons can be obtained by summing over the two transverse acoustic modes and one longitudinal acoustic mode~\citep{Swartz1989}:
\begin{equation}
\Sigma_K=\frac{\pi^2 k_B^4}{120 \hbar^3}\frac{\left(\frac{1}{v^2_{1L}}+\frac{2}{v^2_{1T}} \right) \left(\frac{1}{v^2_{2L}}+\frac{2}{v^2_{2T}} \right)}{\left( \frac{1}{v^2_{1L}}+\frac{2}{v^2_{1T}}+\frac{1}{v^2_{2L}}+\frac{2}{v^2_{2T}} \right)}
\end{equation}
where $v_{iL}$ and $v_{iT}$ are the longitudinal and transverse sound velocities in medium $i$. In the case of solid-solid boundaries, the DMM and the AMM give similar predictions. For simplicity, we use the DMM for estimation of the Kapitza resistance. To calculate $\Sigma_{K}$, we use the values of the sound velocity reported for Ni~\cite{Swartz1989}, amorphous SiO$_2$~\cite{Love1973}, and the cubic phase of NbN~\cite{Ren2012}. $v_{L}$ and $v_{T}$ in InO$_x$ are obtained from elastic constants and the density of In$_2$O$_3$~\cite{Walsh2009,Ellmer2001}. We find that $\Sigma_{K}$ covers the range from $230$ to $300$\,W K$^{-4}$m$^{-2}$ for the thermal contacts of NbN-SiO$_2$, Ni-SiO$_2$, and In$_2$O$_3$-SiO$_2$. \mycom{The thermal impedance mediated by the Kapitza resistance $Z_K$ can be calculated as $Z_{K}=1/G$, where the effective thermal conductance $G=4\Sigma_{K}T^3$. The temperature dependence of $Z_K$ for all interfaces is shown in the inset in \autoref{fig_02}}.

\section*{\label{ap4} Appendix D: Derivation of thermal conductivity}

Integrating the expression for the heat flow
\begin{equation}
\begin{aligned}P_{2D}=\kappa_{AIL}\frac{dT_{ph}}{dz}=const\left(z\right)\end{aligned}
\label{HeatFlow}
\end{equation}
over the thickness of the AIL layer, one gets 
\begin{equation}
\begin{aligned}P_{2D} d_{AIL}=\int_{0}^{d_{AIL}}\kappa_{AIL}\frac{dT_{ph}}{dz}dz\end{aligned}
=\begin{aligned}\int_{T_{b}}^{T_{e}}\kappa_{AIL} dT_{ph}\end{aligned}
,\label{IntHeatFlow}
\end{equation}
which, after differentiation over the electron temperature, yields
\begin{equation}
\frac{dP_{2D}}{dT_{e}}d_{AIL}=\frac{d}{dT_{e}}\int_{T_{b}}^{T_{e}}\kappa_{AIL} dT_{ph}=\kappa_{AIL}\left(T_{e}\right),\label{kappa_T_e}
\end{equation}
that is, Eq.~\ref{eq_kox}.

\section*{\label{ap5} Appendix E: Estimation of SiO$_2$ and Si bottlenecks}
In the main text, the thermal resistance $Z$ of the substrate is determined as $Z \equiv dT_e/d P_{2D}$. The substrate used in set\,1 is made of an amorphous SiO$_2$ layer on bulk crystalline Si with thicknesses of $d_{AIL}$ and $d_{Si}$, respectively. Since the geometry of the metal films satisfies the conditions $w,L>d_{AIL}$ and $w,L<d_{Si}$, a one-dimensional heat outflow into the AIL and a further three-dimensional heat outflow into the Si substrate are expected~\cite{Pop2010}. The thermal resistances mediated by the AIL and the silicon substrate ($Z_{SiO_2}$ and $Z_{Si}$) can be calculated as 
$Z_{SiO_2}=d_{AIL}/\kappa_{SiO_2}$ (see the derivation in Appendix D and $Z_{Si}=\sqrt{Lw}/\kappa_{Si}$~\cite{Pop2010}. Taking the thermal conductivities $\kappa_{SiO_2}\simeq 0.1$\,WK$^{-1}$m$^{-1}$~\cite{Smith1978} and $\kappa_{Si}\simeq10^4$\,WK$^{-1}$m$^{-1}$ at 10\,K~\cite{Ruf2000} and the maximum value of $w\times L=307$\,$\mu$m$^{2}$, we estimate $Z_{SiO_2}=3\times 10^{-6}$\,W$^{-1}$Km$^2$ and $Z_{Si}=10^{-9}$\,W$^{-1}$Km$^2$ at 10\,K.

\bibliography{Diffuson_bottleneck}
\end{document}